\documentclass[aps, prl,twocolumn, amsfonts, amsmath, amssymb, superscriptaddress, float]{revtex4}
\usepackage{graphicx}
\usepackage{color}
\usepackage[colorlinks, linkcolor={blue},
    citecolor={blue}, urlcolor={blue}]{hyperref}
\begin{document}
\title{Energy landscape in two-dimensional Penrose-tiled quasicrystal}

\author{Patrizia Vignolo}\affiliation{Universit\'e Nice Sophia Antipolis, CNRS, Institut non Lin\'eaire de Nice, UMR 7335, Valbonne, France}

\author{Matthieu Bellec}\affiliation{Universit\'e Nice Sophia Antipolis, CNRS, Laboratoire de la Physique de la Mati\`ere Condens\'ee, UMR 7336, Nice, France}
 
\author{Julian B{\"o}hm}\affiliation{Universit\'e Nice Sophia Antipolis, CNRS, Laboratoire de la Physique de la Mati\`ere Condens\'ee, UMR 7336, Nice, France}
 
\author{Abdoulaye Camara}\affiliation{Universit\'e Nice Sophia Antipolis, CNRS, Institut non Lin\'eaire de Nice, UMR 7335, Valbonne, France}

\author{Jean-Marc Gambaudo}
\affiliation{Universit\'e Nice Sophia Antipolis, CNRS, Institut non Lin\'eaire de Nice, UMR 7335, Valbonne, France}

\author{Ulrich Kuhl}\affiliation{Universit\'e Nice Sophia Antipolis, CNRS, Laboratoire de la Physique de la Mati\`ere Condens\'ee, UMR 7336, Nice, France}
 
\author{Fabrice Mortessagne}\affiliation{Universit\'e Nice Sophia Antipolis, CNRS, Laboratoire de la Physique de la Mati\`ere Condens\'ee, UMR 7336, Nice, France}
 
\date{\today}

\maketitle

{\bf Since their spectacular experimental realisation in the early 80's \cite{Shechman1984}, quasicrystals \cite{Bindi2009} have been the subject of very active research, whose domains extend far beyond the scope of solid state physics. In optics, for instance, photonic quasicrystals have attracted strong interest~\cite{Vardeny2013} for their specific behaviour, induced by the particular spectral properties, in light transport~\cite{Hattori1994, Man2005, Levi2011}, plasmonic~\cite{Gopinath2010} and laser action~\cite{Notomi2004}.
Very recently, one of the most salient spectral feature of quasicrystals, namely the gap labelling~\cite{Bellissard1993}, has been observed for a polariton gas confined in a one dimensional quasi-periodic cavity~\cite{Tanese2014}. This experimental result confirms a theory which is now very complete in dimension one~\cite{Suto1989, Luck1989}. In dimension greater than one, the theory is very far from being complete. Furthermore, some intriguing phenomena, like the existence of self-similar eigenmodes can occur \cite{Sutherland1986}. All this makes two dimensional experimental realisations and numerical simulations pertinent and attractive. Here, we report on measurements and energy-scaling analysis of the gap labelling and the spatial intensity distribution of the eigenstates for a microwave Penrose-tiled quasicrystal. Far from being restricted to the microwave system under consideration, our results apply to a more general class of systems.}

\begin{figure}[t]
\centering
\includegraphics[width=\columnwidth]{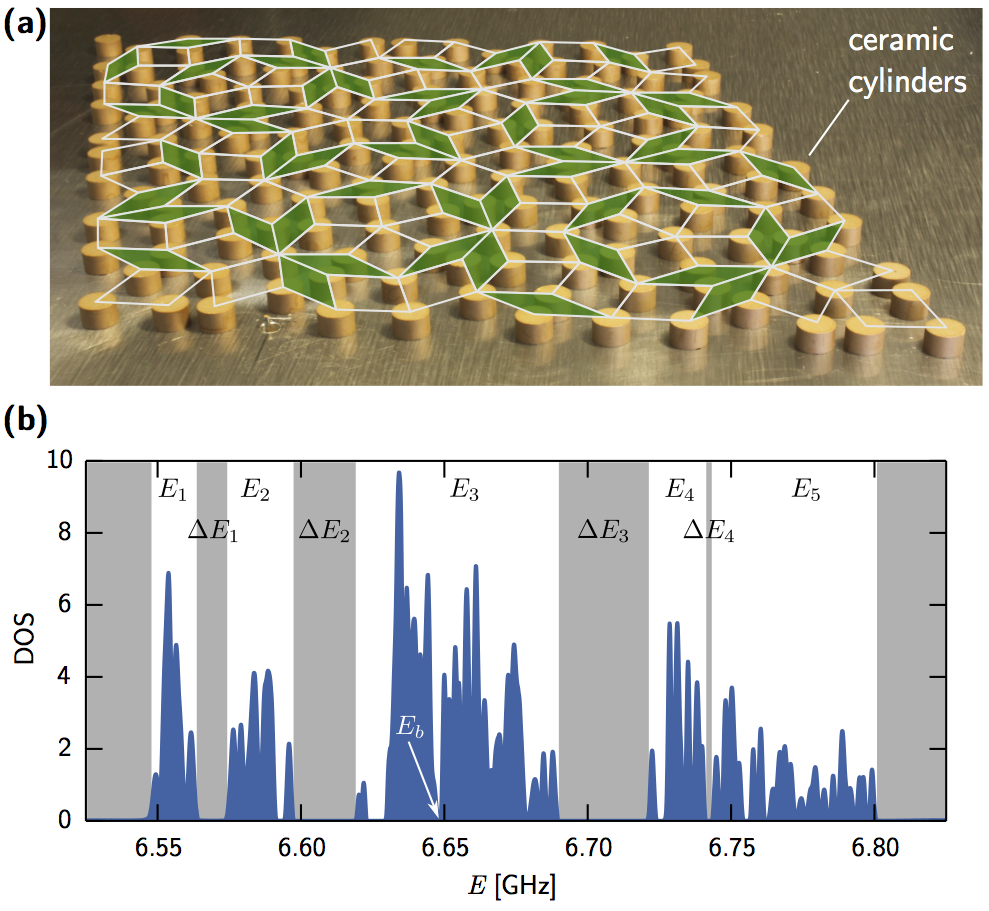}
\caption{\label{fig1} {\bfseries Microwave Penrose-tiled quasicrystal.} \textbf{(a)} Diamond-vertex Penrose-tiled quasicrystal, where the tiling is superposed to guide the eye (thin diamonds in green). The sites of the lattice are occupied by dielectric resonators (ceramic cylinders of 5 mm height and 8 mm diameter) with a high index of refraction ($n=6$). The lattice is sandwiched between two aluminium plates (the upper one is not shown). The microwaves are excited by a movable loop antenna. \textbf{(b)} Experimentally obtained DOS as a function of frequency, the white and gray zones indicate the main frequency bands, $E_i$, and the gaps, $\Delta E_i$, respectively. The bare frequency $E_b$ is indicated by the white arrow.}
\end{figure}

Quasicrystals are alloys that are ordered but lack translational symmetry. In dimension two, they can be modelled with a collection of polygons (tiles) that cover the whole plane, so that each pattern (a sub-collection of tiles) appears, up to translation, with a given density but the tiling is not periodic. 
A typical example is given by the Penrose tiling \cite{Penrose1974}. Here, we implement a microwave realisation of a Penrose-tiled lattice using a set of coupled dielectric resonators [see Fig.~\ref{fig1} (a)]. The microwave setup used has shown its versatility by successfully addressing various physical situations ranging from Anderson localisation~\cite{Laurent2007} to topological phase transition in graphene~\cite{Bellec2013a}, and provided the first experimental realisation  of the Dirac oscillator~\cite{Franco2013}.

We establish a two-dimensional tight-binding regime~\cite{Bellec2013}, where the electromagnetic field is mostly confined within the resonators. 
For an isolated resonator, only a single mode is important in a broad spectral range around the bare frequency $E_b\simeq6.65$\,GHz. 
This mode spreads out evanescently, so that the coupling strength $t$ can be controlled by adjusting the separation distance $d$ between the resonators~\cite{Bellec2013}. 
The resulting system can be described by the following tight-binding Hamiltonian:
\begin{equation}
H=E_b \sum_i |i\rangle\langle i|+\sum_{i,j, i\neq j}t_{ij} |i\rangle\langle j|,
\label{hamilt}
\end{equation}
where $|i\rangle$ is the wave function at site $i$ and $t_{ij}$ is the coupling strength between sites $i$ and $j$. We created a Penrose tiling made of thin and fat diamonds. The lattice is constructed using the inflation rules of the Robinson triangle decomposition \cite{graum1987} of the Penrose tiling \cite{Penrose1974}. A microwave resonator is placed at each diamond vertex [Fig.~\ref{fig1}~(a)]. The experimentally obtained density of states (DOS), $\rho(E)$ (see Supplementary Information for details), is shown in Fig.~\ref{fig1}~(b). The main frequency bands, $E_i$, and gaps, $\Delta E_i$, are indicated by the white and grey zones, respectively. The peak at ``zero-energy", here the bare frequency $E_b$,  predicted when the couplings are restricted to the diamond edges \cite{Kohmoto1986}, is not observed here. Indeed, as explained below, in our system the spectrum is dominated by the couplings along the short diagonal of thin diamonds.

\begin{figure}
\centering
\includegraphics[width=\columnwidth]{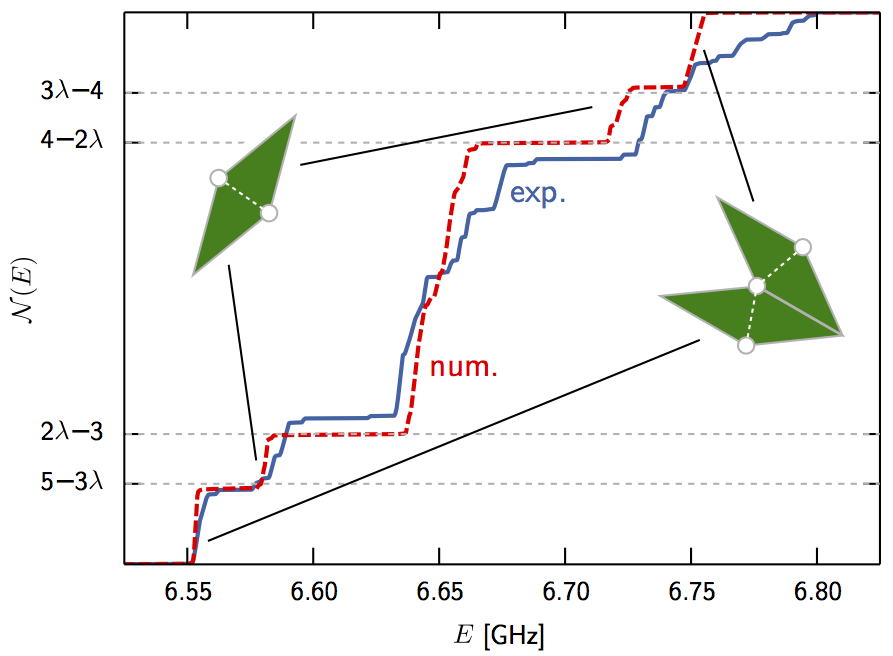}
\caption{\label{fig2} {\bfseries Integrated density of states.} Normalised IDOS $\mathcal{N}(E)$ for a diamond-vertex
Penrose lattice setup. Solid red line corresponds to the experimental Penrose lattice with 164 sites (see Fig.~\ref{fig1}). The dashed blue line is obtained by directly diagonalising the tight-binding Hamiltonian using 2665 sites with the corresponding coupling constants obtained by Eq.~(9) of Ref. \cite{Barkhofen2013}). The horizontal dashed lines indicate the gap labelling for an infinite lattice. Isolated and coupled green diamonds are respectively associated with the dimer and trimer structures of closer sites whose couplings generate the five bands $E_i$ in the DOS and give the main structure to the IDOS staircase.} 
\end{figure}

To visualise the gap labelling, we consider the integrated density of states (IDOS): $\mathcal{N}(E)=\int_{-\infty}^E \rho(E') \, \textrm{d}E'$. As plotted in Fig.~\ref{fig2} (solid blue line), we clearly observe plateaus for specific values of $\mathcal{N}(E)$, where the IDOS is approximately constant. In contrast to a periodic crystal, where the number of states in each band is the same and the IDOS is a regular staircase with identical step heights, in a quasicrystal, due to the fractal structure of its spectrum, the IDOS is an irregular staircase. Gap labelling identifies the position of footsteps where $\mathcal{N}(E) $ is constant and depends only on the geometry of the quasicrystal and not on the physical nature of the coupling between sites~\cite{Bellissard1993,Prunele2002,Benameur2003,Kaminker2003,Bellissard2006}. For instance, for an infinite Fibonacci or Penrose quasicrystal, it was shown that each footstep in the IDOS appears at a precise value given by ${\mathcal N}(E)={z}_{i_1}+\lambda{z}_{i_2}$, where $\lambda=(\sqrt{5}+1)/2$ is the golden ratio and the $z_{i}$'s are relative integers. Moreover, the IDOS builds a devil's staircase:  each step is divided in smaller steps and each smaller step is divided in steps even smaller, leading to a spectrum with a Cantor set structure~\cite{Suto1989, Luck1989}.

For an infinite Penrose-tiled quasicrystal, as explained in the Supplementary Information, the gap labelling theory establishes the following main hierarchy of the gaps~\cite{Prunele2002}:
\begin{equation}
\mathcal{N}(E)= 
\begin{cases}
5-3\lambda, & E \in \Delta E_1 \\
2\lambda-3, & E \in \Delta E_2 \\
4-2\lambda , & E \in \Delta E_3 \\ 
3\lambda-4, & E \in \Delta E_4,
\end{cases}
\label{eqgap}
\end{equation}
$\Delta E_i$ being the $i$th frequency gap. As shown in Fig.~\ref{fig2}, these values perfectly fit for the IDOS calculated numerically for a large system (dotted red line). The experimental IDOS (solid blue line) is in very good agreement with the predicted hierarchy. The discrepancies between the numerics and the experiments have three main reasons. First, the experimental lattice only possesses a few tens of states in each band; the finite-size effect is particularly visible in the last band, which is the largest. Second, the spatial extension of the resonators is not taken into account in the numerics, thus neglecting screening effects. Third, the microwave resonators are not strictly speaking identical: the relative variation of the bare frequency can reach $0.15\%$.

\begin{figure*}
\includegraphics[width=2\columnwidth]{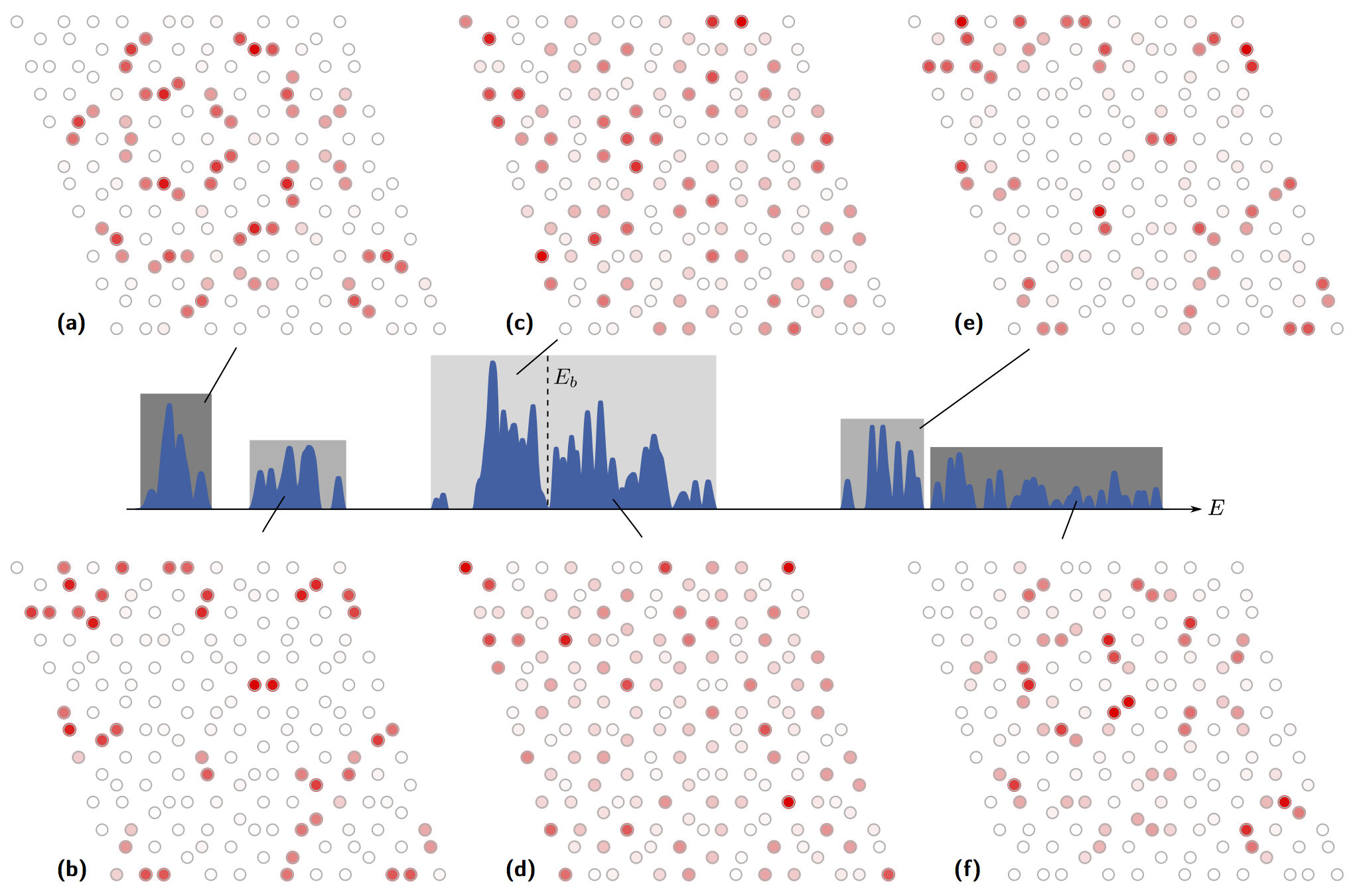}
\caption{\label{fig3} {\bfseries Energy landscape in each band.} The central part of the figure shows the DOS for the diamond-vertex tiling. It exhibits five main bands distributed on both sides of the bare frequency $E_b$ indicated by the vertical dotted line, the central band showing a point-like dip at $E_b$. For each single resonator we integrate the LDOS band by band and obtain the overall wave function structure within the bands. (a) and (f) Energy localised on trimers of closer sites is associated with the two extreme bands (dark grey zones). (b) and (e) For the two neighbouring bands (medium grey zone) a localisation on dimer structures is observed. (c) and (d) For the central band (light grey zone), integrations from the lower limit to the bare frequency and from the bare frequency to the upper limit give similar uniformly distributed energy landscapes: energy is delocalised over the edge sites of the trimer structures and over the ``uncoupled'' resonators.}
\end{figure*}

For a better understanding of the spectrum, we analyse the wave function spatial distribution band by band (Fig.~\ref{fig3}). The data processing from reflexion measurements to the intensity of wave functions, via LDOS, is reported in the Supplementary Information.
We observe that, in the first and last bands (dark grey zones), the energy is mainly distributed in trimers of closer sites, with a maximum of intensity in the central resonator of each trimer [Figs~\ref{fig3}~(a) and (f)]. In the second and fourth bands (medium grey zones), the energy is essentially distributed in dimers of closer sites [Figs~\ref{fig3}~(d) and (c)]. In the central band (light grey zone), we observe that the energy is distributed more uniformly [Figs~\ref{fig3} (b) and (e)]. Moreover, the dimers and trimers that emerge in the landscape are associated with the two patterns corresponding to isolated or coupled thin diamonds depicted in green in Fig.~\ref{fig2}. 

Dimer and trimer couplings give the dominant contribution to the spectral structure. This can be understood following a perturbative approach. If there were no coupling among the resonators, the spectrum would be a peak at $E_b$ with a degeneracy given by the number of sites in the system. The main Hamiltonian terms which contribute to remove this degeneracy are the ones where the coupling is largest, $t_{ij}=t_\textrm{max}$, i.e. those associated with resonators located at the minimum distance $d_\textrm{min}$. This occurs for the resonators located at the closest vertices of the thin diamonds. Thus, in the isolated thin diamonds we have a dimer structure $\{|1_d\rangle,\;|2_d\rangle\}$, while in the coupled diamonds we have a trimer structure $\{|1_t\rangle,\;|2_t\rangle,\;|3_t\rangle\}$. If the other couplings were vanishing, the spectrum would have five peaks located at the energies
\begin{equation}
\begin{split}
&E_1=E_b-\sqrt{2}t_\textrm{max},\;\;\;E_2=E_b-t_\textrm{max},\\
&E_3=E_b,\\
&E_4=E_b+t_\textrm{max},\;\;\;E_5=E_b+\sqrt{2}t_\textrm{max}.
\label{eigenvalues}
\end{split}
\end{equation}
These eigenvalues would correspond to the eigenfunctions
\begin{equation}
\begin{split}
&|\phi_1\rangle=|1_t\rangle-\sqrt{2}|2_t\rangle+|3_t\rangle,\;\;\;|\phi_2\rangle=|1_d\rangle-|2_d\rangle,\\ & |\phi_{3,a}\rangle=|1_t\rangle- |3_t\rangle,\;\;\;|\phi_{3,b}\rangle=|{1_s}\rangle,\\
&|\phi_4\rangle=|1_d\rangle+ |2_d\rangle,\;\;\;|\phi_5\rangle=|1_t\rangle+\sqrt{2}|2_t\rangle+|3_t\rangle.\\
\label{eigenstates}
\end{split}
\end{equation}
Eigenvalues $E_1$ and $E_5$ result from the trimer structure, while $E_2$ and $E_4$ from the dimer one.
$E_3$ results from both the trimer structure and all sites $|{1_s}\rangle$ that are considered uncoupled in this perturbative argument.  
Thus the degeneracy of the first and the fifth peaks is given by the number of coupled thin diamonds present in the tiling, while the one of the second and fourth by the number of isolated thin diamonds. Finally, the degeneracy of the central  peak is given by the number of coupled thin diamonds plus the number of all the remaining resonators that are uncoupled. The other couplings are much smaller than $t_\textrm{max}$ since they decrease almost exponentially with distance~\cite{Bellec2013}. Indeed the coupling corresponding to the side of the diamonds is $\sim t_\textrm{max}/10$ and the others are even smaller. The main effect of these weaker couplings is a broadening of the five peaks into five bands, as observed in the spectrum [see Fig.~\ref{fig1} (c)]. The population of the five bands of the spectrum is just given by the reasoning above. Thus the numbers in Eq.~(\ref{eqgap}) are related to the fraction of isolated and coupled thin diamonds in the Penrose tiling as detailed in the Supplementary Information.

The perturbative approach outlined above can also be used to understand the energy landscape shown in Fig.~\ref{fig3}. The trimer structures in the LDOS that appear in the first and in fifth bands [Figs~\ref{fig3} (a) and (f)] correspond to the states $|\phi_1\rangle$ and $|\phi_5\rangle$, respectively.
The dimerised structures in the LDOS that appear in the second and fourth bands [Figs~\ref{fig3} (b) and (e)] correspond to the states $|\phi_2\rangle$ and $|\phi_4\rangle$, respectively. For the central band [Fig.~\ref{fig3} (c) and (d)] the LDOS corresponds to the states $|\phi_{3,a}\rangle$ and $|\phi_{3,b}\rangle$ and thus it vanishes over the dimers and over the central site of each trimer ($|2_t\rangle$). 

We stress that, for larger systems and with better frequency resolution, it would be possible to observe further hierarchical structure of the bands, namely the fragmentation of each band in sub-bands in connection with the density of larger patterns. This may be possible using ultracold gases \cite{Gambaudo2014} trapped in quasicrystal optical lattices~\cite{Guidoni1999}. Another important remark is that the gap labelling values in Eq.~(\ref{eqgap}) do not depend on the particular values of the hopping energies in Eq.~(\ref{hamilt}). For any (non-interacting) classical or quantum system subject to a potential with the same
spatial patterns as the ones studied in this work, the first hierarchy of the gaps will be given by the values in Eq.~(\ref{eqgap}), the only condition being that the shorter the distance between sites the larger their coupling (the study performed in \cite{He1989} does not meet this criterion). Within this condition, different physical systems will exhibit the same energy landscapes as the ones shown in Fig.~\ref{fig3}. What will change from one system to another are the widths of the gaps and bands. 

\begin{figure}
\begin{center}
\includegraphics[width=\linewidth]{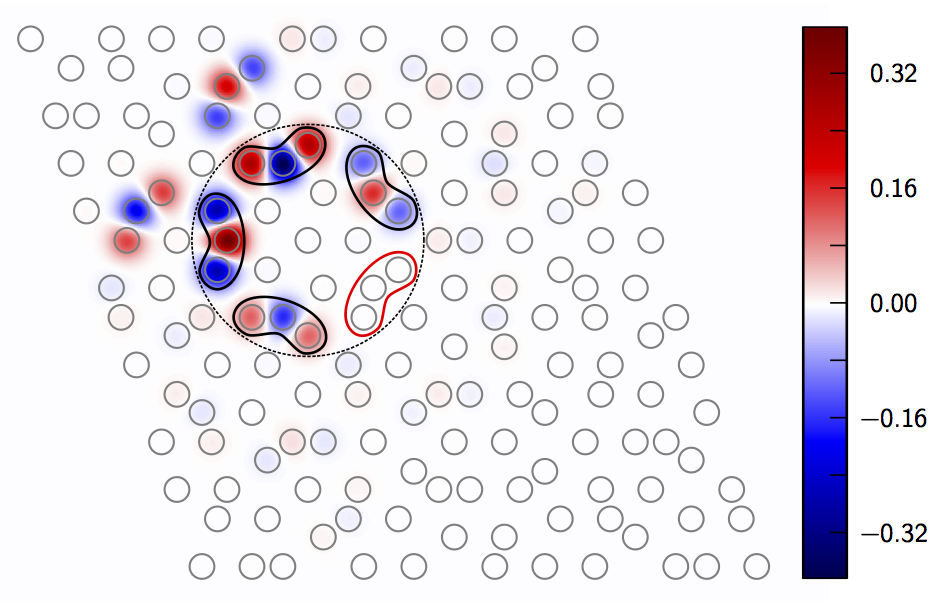}
\caption{\label{fig4} {\bfseries Frustration mechanism.} Ground state wave function (numerical calculation) for the Penrose-tiled quasicrystal. In each trimer the sign alternates between neigbouring sites, and, to minimize the energy, the sign alternates also between neighbouring trimers. The dotted black circle delineates a pattern of five trimers exhibiting a geometrical frustration effect: one of the trimers has to possess a vanishing energy since the complete alternation of signs cannot be fulfilled.}
\end{center}
\end{figure}

The difference between states $|\phi_1\rangle$ and $|\phi_5\rangle$ (and similarly for $|\phi_2\rangle$ and $|\phi_4\rangle$) is the phase landscape. Because of the positive sign of the coupling in such an experiment \cite{Bellec2013}, the ground state corresponds to trimers with alternating signs ($|\phi_1\rangle$). Neighbouring  trimers are also arranged such that the alternance of signs is fulfilled, thus allowing the minimisation of the energy. The main consequences are twofold: (i) the state is almost localised, as outlined in~\cite{Kohmoto1986,Sutherland86PRB}, and indeed the first band is very narrow; (ii) there is a geometrical frustration effect as shown in Fig.~\ref{fig4}: where there are patterns with five trimers forming a circular structure, the wave function is vanishing on one of the trimers (the bottom right trimer of the structure, in Fig.~\ref{fig4}), because it is not possible for all five trimers to satisfy alternate signs with neighbouring trimers. The wave function is numerically obtained by diagonalising the tight-binding Hamiltonian corresponding to the experimental situation (same number of sites and same couplings). The reflexion measurements only provide the intensity of wave functions, the sign is experimentally accessible via transmission measurements~\cite{arXpol14} which were not implemented in this study.

In conclusion we have measured experimentally the gap labelling and the energy landscape of the eigenstates of a microwave Penrose-tiled quasicrystal. To the best of our knowledge, this is the first measure of these two observables in dimension two. In such a microwave system, the Hamiltonian terms scale exponentially with the distance. We have shown that this allows to use a perturbative argument to understand and determine both the gap labelling and the energy landscape of the different states. Even though the derivation of our results is related to our particular setup, the gap labelling values and the wave function symmetry in each band that we have found, are general. Indeed, our results concern all systems which share the same quasicrystalline potential energy landscape. 


\vspace{1cm}
\noindent{\bf Author contributions} M.B., U.K. and F.M. designed the experiment;
M.B., J.B., A.C. and F.M. performed the experiments and analysed the data; P.V. and A.C. performed the numerical simulations; J-M.G. and P.V. calculated the gap labelling; all the authors participated to the discussion of the results and the writing of the paper.

\bigskip
\noindent{\bf Correspondence} Correspondence and requests for materials should be addressed to P.V. (email: Patrizia.Vignolo@inln.cnrs.fr) and F. M. (email: Fabrice.Mortessagne@unice.fr).

\clearpage

\appendix

\makeatletter
\renewcommand{\theequation}{S.\arabic{equation}}
\renewcommand{\thefigure}{S\@arabic\c@figure}
\makeatother
\setcounter{figure}{0}
\setcounter{equation}{0}

\section*{SUPPLEMENTARY INFORMATION}

\noindent \textbf{S1. DOS and wavefunctions extractions}

\medskip
The experimental setup and the tight-binding description of the microwave system are detailed in~\cite{Bellec2013}. Here, we describe the procedure used to obtain the density of states (DOS) and the wavefunctions from reflection measurements. 

\begin{figure}[h]
\begin{center}
\includegraphics[width=\linewidth]{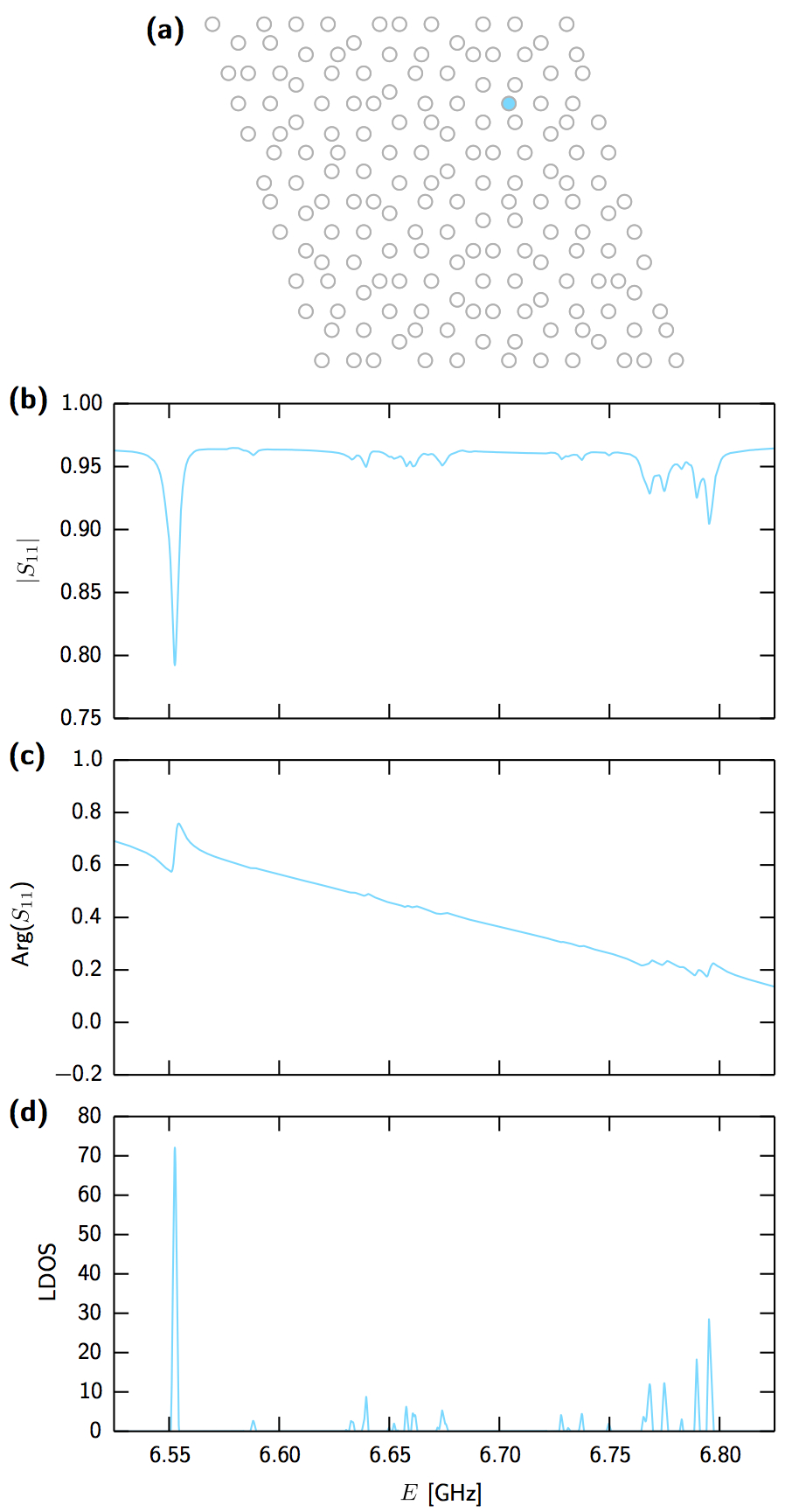}
\caption{{\bfseries From reflexion signal to LDOS.} \textbf{(a)} Penrose-tiled quasycristal whose sites are occupied by microwave resonators. The coloured site (light blue) indicates the position of the magnetic antenna connected, via a coaxial cable, to a vectorial network analyser.  \textbf{(b)} and \textbf{(c)} The corresponding amplitude and phase of the reflexion signal $S_{11}$. \textbf{(d)} The local density of states deduced from the reflexion measurement according to relation (\ref{Eq_rho}).}
\label{Fig_Supp_Spectra} 
\end{center}
\end{figure}

The resonance frequency of an isolated resonator $E_b$ is around $6.65$\,GHz and is associated with the first Mie resonance in the transverse electric polarisation. Due to evanescent coupling, a lattice of such resonators exhibits a discrete spectrum a few hundreds of MHz wide around $E_b$. Figure~\ref{Fig_Supp_Spectra} shows the reflection signal $S_{11}$ measured with a vectorial network analyser, via a magnetic antenna positioned above the centre of a given resonator in a Penrose-tiled quasycristal.
We have shown in~\cite{Bellec2013} that the local density of states (LDOS), $\rho(\mathbf{r}_1, E)$, measured at position $\mathbf{r}_1$, is directly related to the amplitude and phase of $S_{11}$. Up to a normalisation factor, the relation reads:
\begin{equation}
\rho(\mathbf{r}_1, E) \propto \frac{\vert S_{11}(E)\vert^2}{\langle \vert S_{11}\vert^2 \rangle_{E}} \varphi_{11}'(E)
\label{Eq_rho}
\end{equation}
where $\langle\ldots\rangle_E$ indicates an averaging over the whole range of the frequency spectrum, $\varphi_{11}$ is the phase of the reflected signal: $\varphi_{11}=\mathrm{Arg}(S_{11})$, and where $\varphi_{11}'$ denotes its derivative with respect to the frequency.

\begin{figure}[h]
\begin{center}
\includegraphics[width=\linewidth]{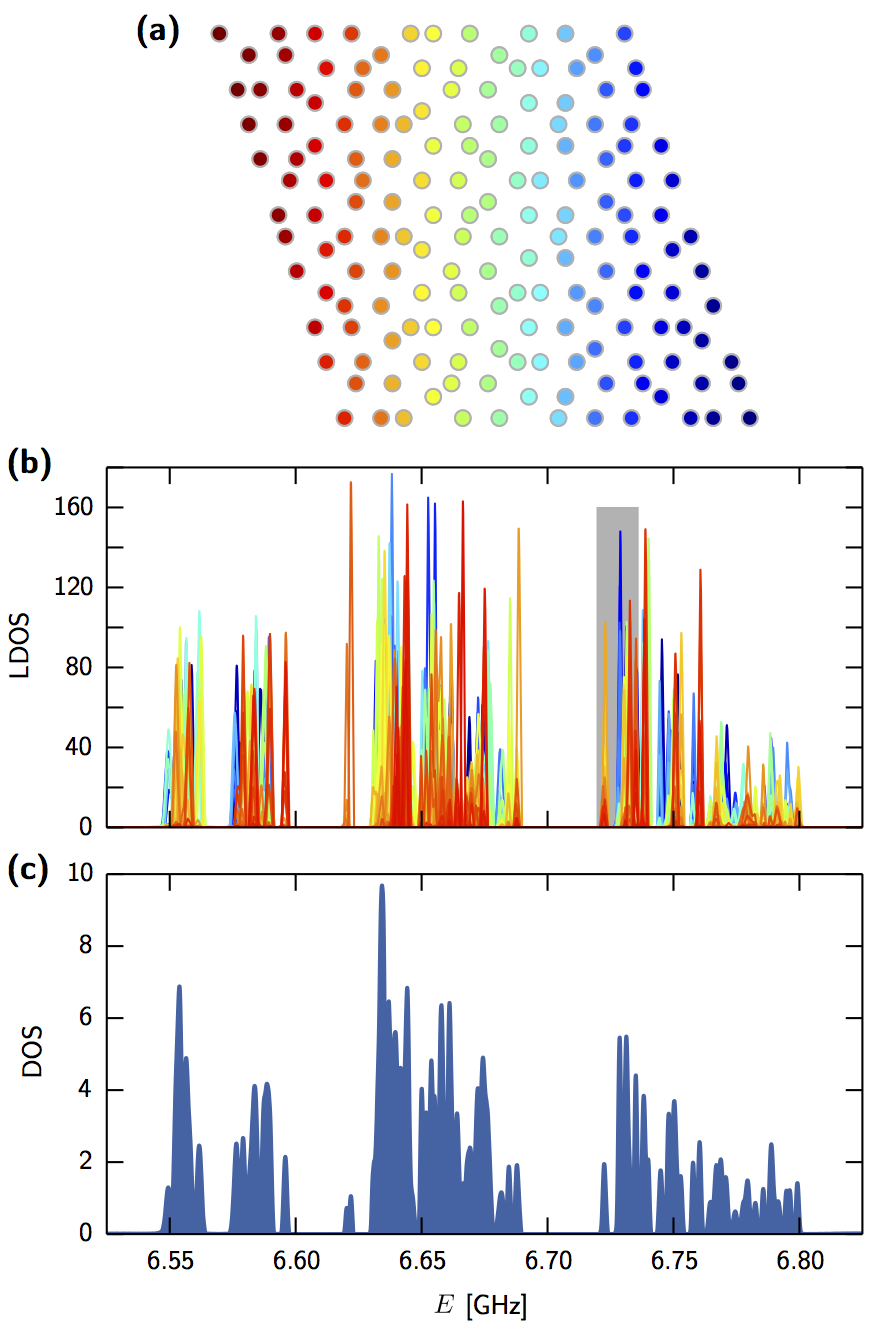}
\caption{{\bfseries From LDOS to DOS.} \textbf{(a)} Color representation of site positions and \textbf{(b)} the corresponding LDOS. \textbf{(c)} Density of states obtained by averaging the LDOS over all site positions.}
\label{Fig_Supp_DOS} 
\end{center}
\end{figure}

In Fig.~\ref{Fig_Supp_DOS}(a), each color (from deep blue to red) corresponds to a given site position. Their respective LDOS are plotted in Fig.~\ref{Fig_Supp_DOS}(b). The density of states (DOS) shown in Fig.~\ref{Fig_Supp_DOS}(c) is obtained by averaging the LDOS over all positions.

\begin{figure}[h]
\begin{center}
\includegraphics[width=\linewidth]{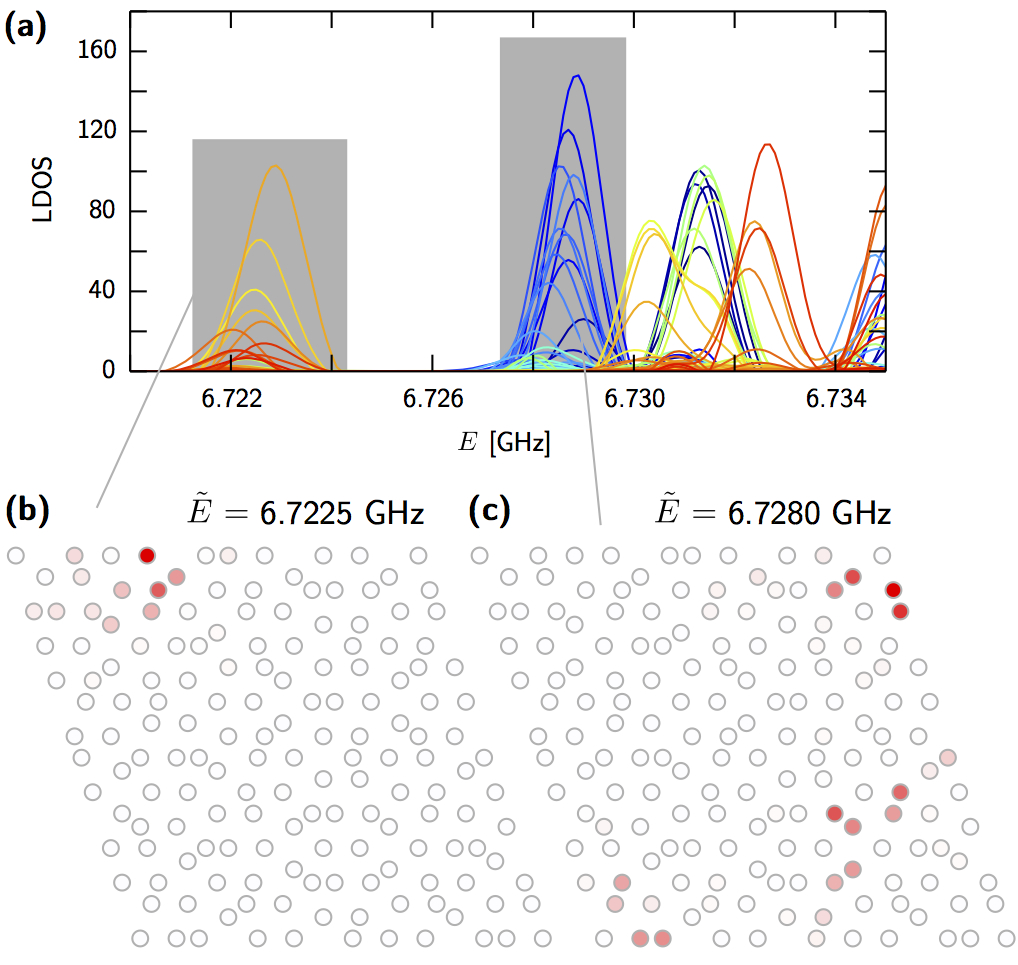}
\caption{{\bfseries From LDOS to wave functions.} \textbf{(a)} A zoom of the LDOS shown in Fig.~\ref{Fig_Supp_DOS}. \textbf{(b)} The wave function intensity distributions respectively associated with the two resonances delineated by the grey boxes. For a given resonance, the intensity on each site is given by the maximum of the corresponding LDOS in the vicinity of the mean eigenfrequency $\tilde{E}$.}
\label{Fig_Supp_WF} 
\end{center}
\end{figure}

Figure~\ref{Fig_Supp_WF}(a) presents a magnification of the small frequency range indicated by the grey zone in Fig.~\ref{Fig_Supp_DOS}(b). For a given eigenfrequency (grey boxes) and a given position, according to the definition of the local density of states $\rho(\mathbf{r}_1, E) =\sum_n\vert\Psi_n(\mathbf{r}_1)\vert^2\delta(E-E_n)$, the resonance curve exhibits a maximum whose value is related to the intensity of the wavefunction at this specific position. The visualisation of the wavefunction distribution associated with each eigenfrequency thus becomes accessible, see Figs~\ref{Fig_Supp_WF}(b) and (c). The energy landscapes depicted in Fig.~\ref{fig3} are obtained by superimposing the wavefunction intensities of all the resonances in the same band.

\begin{figure}[h]
\begin{center}
\includegraphics[width=0.9\linewidth]{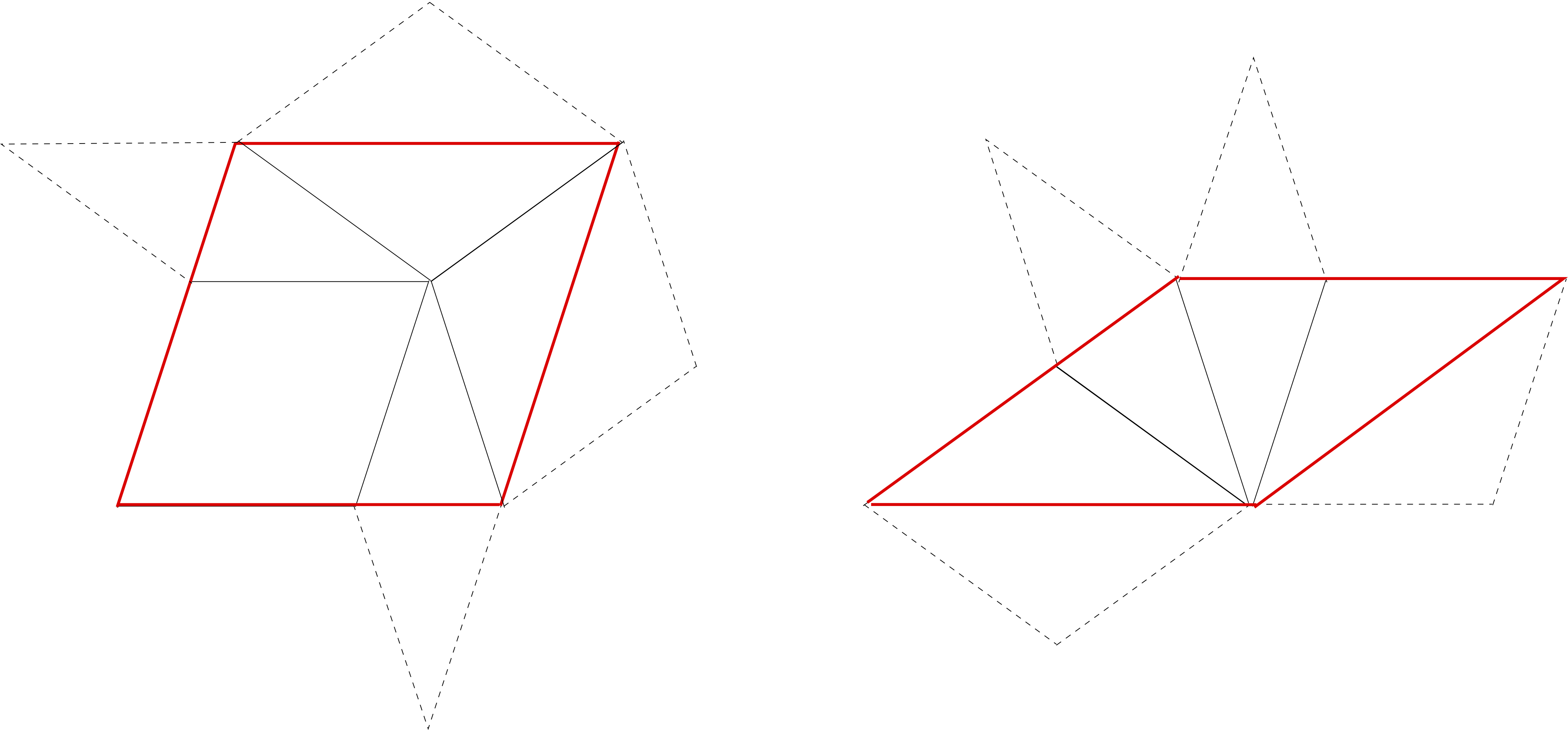}
\caption{\label{Fig_Supp_losange} Inflation of the basic tiles (one iteration) to produce a larger pattern. The dashed lines are a guide to the eye to identify the tiles which appear in the new pattern.}
\end{center}
\end{figure}

\clearpage
\noindent \textbf{S2. Calculation of the gap labelling}
In Fig.~\ref{fig2} and in Eq.~(\ref{eqgap}) we have given the values of the IDOS that label the main gaps of the system.
These values can be deduced by analysing the inflation transformation used to generate the diamond Penrose tiling,
as the one shown in Fig.~\ref{Fig_Supp_losange}.
For instance, for such an experiment, we started with a fat diamond and we iterated the inflation procedure 10 times (16 times for the theoretical curves shown in Fig.~\ref{fig2}). At the $i$-th iteration, the number of fat diamonds $N_f^i$ and thin diamonds $N_t^i$ is related by the numbers at the preceding iteration by the expression
\begin{equation}
\left(\begin{matrix}N_f^i\\N_t^i\end{matrix}\right)=
\mathcal{A}
\left(\begin{matrix}N_f^{i-1}\\N_t^{i-1}\end{matrix}\right)
\;\;\;{\rm with}\;\;\;
\mathcal{A}=\left(\begin{matrix}2\phantom{bl} & 1\\1\phantom{bl}&1\end{matrix}\right).
\end{equation}
The eigenvalues of $\mathcal{A}$ are
$\lambda_+=\lambda+1$ and $\lambda_-=(\lambda+1)^{-1}$,
and the corresponding eigenvectors ${\bf u}_+$ and ${\bf u}_-$ have components $(\lambda,1)$ and $(1,-\lambda)$. The eigenvector corresponding to the inflation is ${\bf u}_+$, so that the ratio $u_{+,1}/u_{+,2}$ can be identified with the ratio $N_f^i/N_t^i$ for large values of $i$. Thus
\begin{equation}
\lim_{i\rightarrow\infty} \dfrac{N_f^i}{N_t^i}=\dfrac{u_{+,1}}{u_{+,2}}=\lambda.
\end{equation}
We remark that, always in the limit of large $i$, the number of vertices (sites) $N_v^i$ can be identified with the total number of tiles $N_{tot}^i=N_f^i+N_t^i$.  Indeed, $N_{tot}^i-N_r^i+N_v^i=1$, where $N_r^i$ is the number of ribs, that for a diamond tiling is equal to twice the number of tiles, $N_r^i= 2N_{tot}^i$. This leads to $N_v^i\simeq N_{tot}^i $.

By looking at Fig.~\ref{figsuppl}, one can notice that, after a single inflation, a fat diamond can never give coupled thin diamonds, while the inflation of a thin diamond does. Thus the number $N_c^i$ of coupled thin diamonds at the $i$-th iteration is equal at the number $N_t^{i-1}$ of thin diamonds at the preceding iteration.

Thus we can write the identity
\begin{equation}
N_{tot}^i=3N_f^{i-1}+2N_t^{i-1}=(3\lambda+2)N_t^{i-1}=(3\lambda+2)N_c^{i}
\end{equation}
from which we can deduce the population $\beta_1$ and $\beta_5$ of the first and last bands in the limit of an infinite system:
\begin{equation}
\beta_1=\beta_5=\dfrac{N_c^i}{N_v^i}=
(3\lambda+2)^{-1}=5-3\lambda.
\end{equation}
The populations $\beta_2$ and $\beta_4$ of the second and fourth bands are given by the ratio $N_{is}^i/N_v^i$, $N_{is}^i$ being the number of
isolated thin diamonds. Taking into account that $N_t^i=2N_c^i+N_{is}^i$, we get
\begin{equation}
\beta_2=\beta_4=\dfrac{N_{is}^i}{N_v^i}=(\lambda+1)^{-1}-2(5-3\lambda)=5\lambda-8.
\end{equation}
The population $\beta_3$ of the third band will be $1-2\beta_1-2\beta_2=7-4\lambda$. The 4 values of the gap labelling shown in Fig.~\ref{fig2} correspond to: $\mathcal{N}_1(E)=\beta_1$, $\mathcal{N}_2(E)=\beta_1+\beta_2$, $\mathcal{N}_3(E)=\beta_1+\beta_2+\beta_3$ and $\mathcal{N}_4(E)=\beta_1+\beta_2+\beta_3+\beta_4$.

\end{document}